\documentclass[aps,prl,twocolumn,superscriptaddress,showpacs,showkeys,amsmath,amssymb]{revtex4}
\usepackage{graphicx}
\usepackage{dcolumn}
\usepackage{bm}
\usepackage{color}
\usepackage[thinspace,thinqspace,amssymb,pstricks]{SIunits}
\usepackage{longtable}
\usepackage[letterpaper,hypertex]{hyperref}
\hypersetup{
    pdftitle={Strong 3p-T$_{1u}$ Hybridization of Ar@C$_{60}$},
    pdfauthor={Martin Morscher, Ari Seitsonen, S. Ito, H. Takagi, Nita Dragoe, Thomas Greber},
    pdfsubject={State of endohedral Ar in C$_{60}$.}, pdfkeywords={Endofullerenes, Hybridization, Photoemission, Photoelectron Diffraction, Photoemission cross section}}

\begin{document}

\title{Strong 3p -T$_{1u}$ Hybridization in Ar@C$_{60}$}
\date{\today}

\author{M. Morscher}
\affiliation{Physik-Institut, Universit\"{a}t Z\"{u}rich, Winterthurerstrasse 190, CH-8057 Z\"{u}rich, Switzerland}
\author{A. Seitsonen}
\affiliation{Physikalisch-Chemisches Institut, Universit\"{a}t Z\"{u}rich, Winterthurerstrasse 190, CH-8057 Z\"{u}rich, Switzerland, and IMPMC, CNRS and Universit\'{e} Pierre et Marie Curie, 4 place Jussieu, case 115, F-75015 Paris Cedex 05, France}
\author{S. Ito}
\affiliation{Department of Advanced Materials, University of Tokyo, Kashiwa-no-ha 5-1-5, Kashiwa 277-8561, Japan}
\author{H. Takagi}
\affiliation{Department of Advanced Materials, University of Tokyo, Kashiwa-no-ha 5-1-5, Kashiwa 277-8561, Japan}
\author{N. Dragoe}
\affiliation{Laboratoire d'\`Etude des Mat\'eriaux Hors \'Equilibre, (LEMHE-ICMMO), UMR 8182- 
CNRS, Universit\'e Paris Sud, F-91405 Orsay, France }
\author{T. Greber}
\email{greber@physik.uzh.ch}
\affiliation{Physik-Institut, Universit\"{a}t Z\"{u}rich, Winterthurerstrasse 190, CH-8057 Z\"{u}rich, Switzerland}

\begin{abstract}
Multilayers of fullerenes with and without endohedral Ar units, C$_{60}$ and Ar@C$_{60}$, were investigated by photoemission and density functional theory. The stoichiometry and the endohedral nature of Ar is checked by x-ray photoelectron spectroscopy and x-ray photoelectron diffraction. Valence band ultraviolet photoemission spectra show a strong hybridisation of the Ar 3p valence shell with the 6T$_{1u}$  molecular orbital of C$_{60}$. A hybridisation gap of 1.6 $\pm$ 0.2 eV is found. This is in agreement with density functional theory (DFT) that predicts 1.47 eV, and indicates Ar@C$_{60}$ to be a noble gas compound  with a strong coupling between Ar and the C$_{60}$ cage. No giant Ar photoemission cross section as predicted for the gas phase in [Phys. Rev. Lett. {\bf 99} 243003 (2007)] was found.
\end{abstract} 

\pacs{71.20.Tx,33.60.+q,79.60.Dp,61.05.js}

\keywords{Endofullerenes, Hybridisation, Photoemission, Photoelectron diffraction, Photoemission cross section}

\maketitle

Shortly after the discovery of C$_{60}$ \cite{kro85}, it was proposed that fullerene carbon cages could be filled with other atoms or molecules \cite{hea85}. 
The realization of such molecules, called endofullerenes or incar-fullerenes was expected to lead to new functionalities,  where the endohedral units are isolated by the carbon cage from the surrounding.
Single nitrogen atoms in C$_{60}$ are a prominent example  \cite{mur96}, where the paramagnetic nature of atomic nitrogen even lead to the idea to use N@C$_{60}$ as a Q-bit \cite{har02}. 

Nuclear magnetic resonance \cite{sau94} and electron spin resonance  \cite{mur96} were the first probes of the interior of fullerenes, and photoemission allowed the determination of the valency of endohedral units \cite{pic97}.
The first view inside endofullerenes came from spectacular transmission electron microscopy experiments on so called peapods, where single Gd atoms were seen inside C$_{82}$, which were lined up in a single wall nanotube \cite{sue00}. Only recently, x-ray photoelectron diffraction allowed a direct look on the arrangement of Dy$_3$N in C$_{80}$ \cite{tre09}.

Fullerenes containing noble gases were particularly useful for studies on the influence of the endohedral unit on the molecular properties \cite{sau94,sau96}.
For Ar@C$_{60}$ it was e.g. shown that in K$_3$Ar@C$_{60}$ samples the superconducting transition temperature decreased compared to K$_3$C$_{60}$  \cite{tak06}. Also it was predicted that the dynamic coupling between Ar and the C$_{60}$ cage would lead, near the C$_{60}$ plasmon frequency, to a giant photoemission cross section enhancement  \cite{mad07}.

All these phenomena call for a better understanding of the coupling between the endohedral unit and the fullerene cage.
In this letter we explore Ar@C$_{60}$ layers by means of photoemission, where a comparison with C$_{60}$ allows the quantitative determination of the hybridisation between Ar and C$_{60}$. The hybridisation turns out to be larger than the Ar valence band width in condensed Ar, which establishes Ar@C$_{60}$ to be a noble gas compound.

Photoemission experiments rely on highly purified samples. For endohedral fullerenes, the synthesis is difficult due to the low production yield and the many purification cycles by High-Pressure Liquid Chromatography (HPLC). Several milligrams of Ar@$C_{60}$ have been produced with a purity $>$95\% \cite{tak06}. To efficiently deposit the molecules on a substrate, we employed a custom-made evaporator with mini Knudsen cells that can be closely approached to the sample ($\sim$ 2-3 cm). This allows the preparation of layers from small amounts of material. We used about 10 $\mu$g of Ar@C$_{60}$.
The experiments were performed in a modified VG ESCALAB 220 photoemission spectrometer  with a base pressure of $<5\cdot10^{-10}$ mbar \cite{gre971}. All data were measured at room temperature. As a substrate we used an Al(111) single crystal that was cleaned by repeated cycles of neon ion sputtering (15 min., 1 keV, $\sim$ 1,5 $\mu A/cm^2$) and annealing to $\sim$ 700 K. The coverages and the cleanliness of the samples were examined with x-ray photoelectron spectroscopy (XPS). 
The molecular ordering and the endohedral position of argon were evidenced by x-ray photoelectron diffraction (XPD) \cite{ost91}. Valence band photoemission spectra were recorded with monochromatized {HeÊI$\alpha$} radiation ($\hbar\omega=21.2$ eV).
Experiments with layers between 3 and 7 monolayers of C$_{60}$ or Ar@C$_{60}$ have been performed.

The gas-phase geometric  and electronic structure of Ar@C$_{60}$ and C$_{60}$ was determined 
using density functional theory (DFT) and the wave function-based MP2 method with the computer code TurboMole \cite{ahl89}. Gaussian basis set TZVPP \cite{sch94} was used in both calculations, and the exchange-correlation functional employed in the DFT calculations was the hybrid PBE0 \cite{ada99}. 

\begin{figure}
	\centering
		\includegraphics[width=0.9\columnwidth]{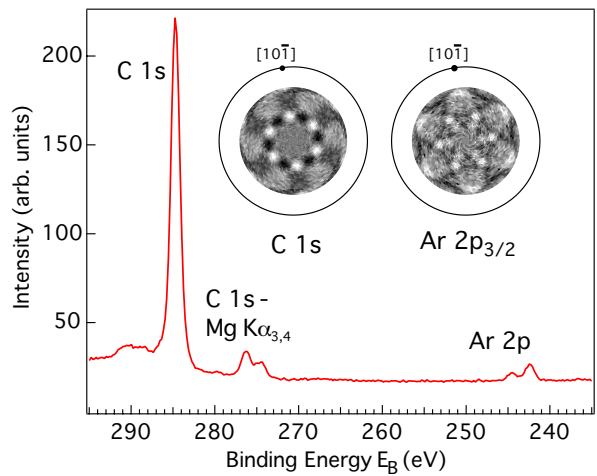}
	\caption{Mg K$\alpha$ X-ray photoelectron spectrum (XPS) and corresponding angle scanned x-ray photoelectron diffraction (XPD) pattern of  Ar@C$_{60}$. XPS indicates a film thickness of 7 monolayers and a C:Ar stoichiometry of 63$\pm$2:1. The C 1s ($E_B$=284.7 eV) and Ar 2p$_{3/2}$ ($E_B$=242.4 eV)  XPD patterns show azimuthal ordering of the molecules, where the high anisotropy ratio between C 1s and Ar 2p of 6.8  indicates that Ar sits inside the carbon cages.}
	\label{F1}
\end{figure}

Figure \ref{F1} shows the characterisation of an Ar@C$_{60}$ layer on Al(111).
The x-ray photoelectron spectrum consists in a dominant C 1s and weak Ar 2p peaks. From the intensity ratio and the atomic cross sections a C:Ar atomic ratio of  56 $\pm7$ :1 is inferred from two different preparations. This is consistent with the nominal stoichiometry of Ar@C$_{60}$ and indicates no significant contribution of contaminations containing carbon, like e.g. C$_{60}$ molecules from an incomplete purification process. 
In contrast to early reports \cite{dic96}, no evidence for depletion of Argon was found under Mg K$\alpha$ and He I$\alpha$ radiation. 
Like for Dy$_3$N@C$_{80}$ on Cu(111) \cite{tre09} the XPD patterns in Figure \ref{F1} have six fold rotational symmetry for the carbon cage as well as for the endohedral unit. The fact that the anisotropy of the Ar signal is 6.8 times larger than that of the carbon pattern is in line with Ref.~\cite{tre09} and evidences that Ar sits inside the C$_{60}$ cage.

\begin{figure}
	\centering
		\includegraphics[width=0.9\columnwidth]{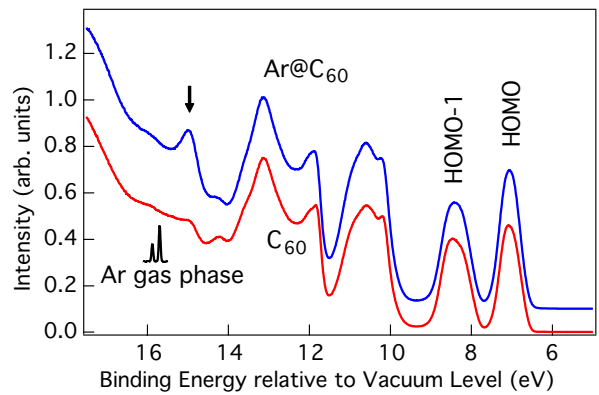}
	\caption{(Color online) He I$\alpha$ excited normal emission  spectra of Ar@C$_{60}$ (blue), C$_{60}$ (red) and gas phase Ar (black). The energies are referred to the vacuum level. 
The arrow at 15 eV indicates the Ar peak in the Ar@C$_{60}$ spectrum, which lies, due better screening of the photoemission final state, above the Ar 3p gas phase lines.} 
	\label{F2}
\end{figure}

Figure \ref{F2} shows the valence band photoemission spectra of multilayers of C$_{60}$ and Ar@C$_{60}$. The two spectra look similar and are dominated by the molecular orbitals of the C$_{60}$ cages. The energies are referred to the vacuum level and no significant energy shift between the highest occupied molecular orbital (HOMO) of C$_{60}$ and Ar@C$_{60}$ is observed. However, at about 15 eV Ar@C$_{60}$ has a clear additional feature. As the inset shows, the energy is close to the Ar 3p levels in the gas phase with an ionisation potential of 15.76 eV. 
The lower binding energy of the endohedral Ar is in line with Ref.~\cite{jac88} and photoemission from Ar clusters \cite{her09} and indicates  a better screening of the photoemission final state with respect to atomic Ar.
The argon peak with a width of 0.53 eV is much broader than the Ar 3p gas phase peaks where the spin orbit splitting of 177 meV \cite{her09} is resolved. 
The width is in the order of the dispersion of Ar monolayers on Pb(111), where a value of about 400 meV has been found by Jacobi \cite{jac88}.
There is also an indirect indication on the endohedral species: 
The partial cross section ratio between the two molecular orbitals HOMO and the HOMO-1 is 0.96$\pm0.02$ and 0.84$\pm0.01$ for Ar@C$_{60}$ and C$_{60}$, respectively. 
In view of the known oscillations of the partial photoemission cross sections \cite{ben91,lie95} and its understanding \cite{rud02}, this is an indication that the potential of the endohedral unit influences the phase of photoelectrons from different molecular orbitals differently.
The intensity of the Ar induced feature does however not confirm a giant photoemission cross section as predicted by theory, where it was argued that the coupling of the photon to the C$_{60}$ and the Ar  cage could enhance the cross section due to resonant interchannel coupling between the Ar 3p and the C$_{60}$ photoemission channels \cite{mad07}.

\begin{figure}
	\centering
		\includegraphics[width=0.9\columnwidth]{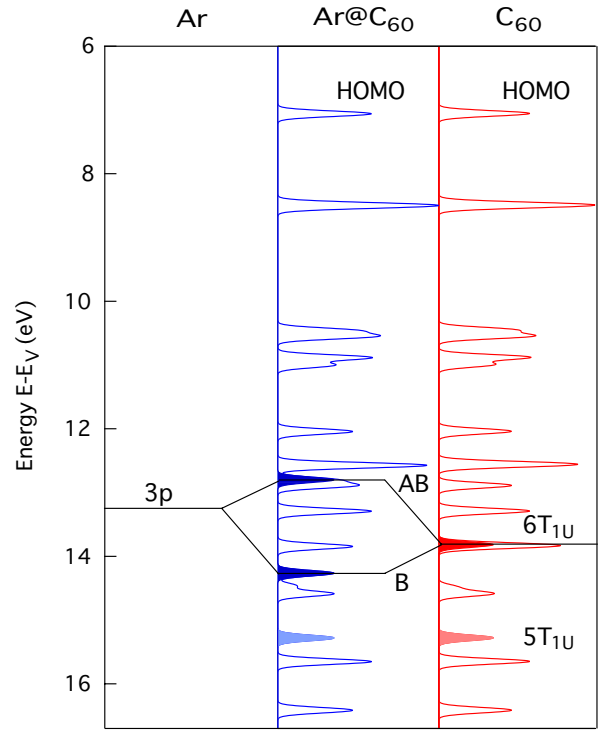}
	\caption{(Color online) Energy eigenvalues (PBE0) of molecular orbitals of C$_{60}$ and  Ar@$C_{60}$ as calculated with density functional theory. The orbital energies are referred to the vacuum level $E_V$ and broadened by a Gaussian with 100 meV full width at half maximum. The orbitals with T$_{1u}$ symmetry are solid, where in C$_{60}$ 5T$_{1u}$ is a $\sigma$- and 6T$_{1u}$ a $\pi$-orbital. 6T$_{1u}$ hybridizes with the Ar 3p shell into a bonding (B) and an antibonding (AB) orbital with a theoretical splitting of 1.47 eV.}
	\label{F3}
\end{figure}

In order to better understand the coupling between the endohedral unit and the C$_{60}$ cage we performed density functional theory calculations that yield the eigenvalues and symmetries of the C$_{60}$ and Ar@C$_{60}$ molecular orbitals. The expectation that the Ar 3p level only interacts with molecular orbitals with the corresponding symmetry (T$_{1u}$) with similar energy and overlap, is nicely confirmed.  
Figure \ref{F3} shows calculated PBE0 eigenvalues of C$_{60}$ and Ar@C$_{60}$. Up to the 6T$_{1u}$ with the same symmetry as the Ar 3p level,  the C$_{60}$ orbitals are unaffected by Ar, i.e. have energy differences for C$_{60}$ and Ar@C$_{60}$ below 25 meV. In C$_{60}$, 5T$_{1u}$ is an orbital with $\sigma$ bond character and  shows no hybridisation (less than 1 meV) due to the lack of overlap.
The 6T$_{1u}$ orbital with $\pi$ character and the nearby Ar 3p orbital hybridize in Ar@C$_{60}$ into a bonding orbital B and an antibonding orbital AB, split by 1.47 eV.
This indicates a strong hybridisation between the endohedral Ar unit and the C$_{60}$ cage.
The 2A$_g$ orbital of C$_{60}$ at an energy of 27.62 eV is not  shown in Figure \ref{F3}. Theory predicts a 455 meV 3s-2A$_g$ hybridisation, though these energy levels are experimentally not accessible with {He I$\alpha$} radiation.

\begin{figure}
	\centering
		\includegraphics[width=0.9\columnwidth]{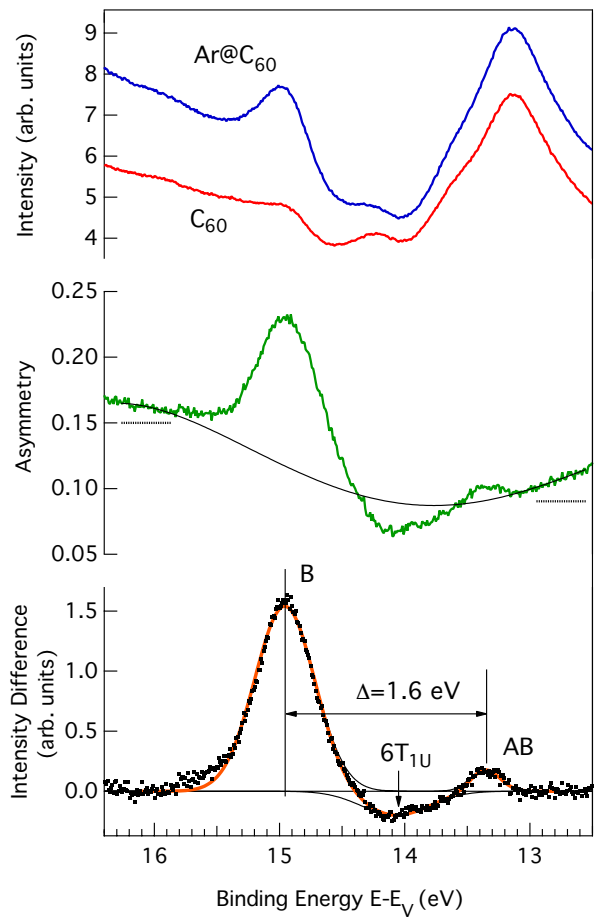}
	\caption{(Color online) Experimental evidence for the 3p -T$_{1u}$ hybridization in Ar@$C_{60}$ (blue) by comparison with C$_{60}$ (red). (a) Raw data as extracted from the spectra in Figure \ref{F2}.(b) Asymmetry between Ar@C$_{60}$ and C$_{60}$ (green) and the background that has been subtracted for quantification. The dashed horizontal lines are the supporting points of the background polynomial. (c) Difference between Ar@C$_{60}$ and C$_{60}$ from (a) and the asymmetry in (b) without background. The splitting $\Delta$ between the bonding and the antibonding hybrid is 1.6 $\pm$  0.2 eV. The negative part of the difference indicates the 6T$_{1u}$ orbital of empty C$_{60}$.}
\label{F4}	
\end{figure}

If we want to compare the theoretical prediction with the experiment we first have to assign the Ar peak (see Figure \ref{F2}) to the bonding or the antibonding orbital.
For this purpose the theoretical molecular orbital eigenvalues are correlated with the experimentally observed molecular orbital peaks \cite{col01}. 
If we assume the deviation betweeen theory and experiment to be proportional to the energy \cite{col01} the PBE0 results suggest that the  experimental Ar peak is 0.64 eV stronger bound than the calculated bonding orbital (B).
This difference between experiment and theory is 2.16 eV, when we assign the antibonding orbital (AB) to the Ar peak at 14.95 eV. For MP2 calculations B also fits with a corresponding difference of -0.57 eV better to the  experiment than AB where the difference is 1.25 eV. 
We therefore assign the experimentally distinct Ar peak to the Ar 3p - C$_{60}$ 6T$_{1u}$ bonding hybrid B.
For the experiment this means that the antibonding 3p-6T$_{1u}$ hybrid orbital must have a lower binding energy than the Ar peak, and that the 6T$_{1u}$ orbital of C$_{60}$ must lie in between them.
A closer inspection of the spectra in Figure \ref{F2} shows that this is the case.
The corresponding  region of interest is shown in Figure \ref{F4}a).
In order to quantify the difference between the two spectra we show the asymmetry A=(I(Ar@C$_{60}$)-I(C$_{60}$))/(I(Ar@C$_{60}$)+I(C$_{60}$)) between the Ar@C$_{60}$ and the C$_{60}$ spectrum in Figure \ref{F4}b).
Clearly, the Ar peak (B) has the largest asymmetry, and 1.6 eV above this main Ar line a new peak shows up. Between the two Ar@C$_{60}$ peaks a C$_{60}$ peak (with a local asymmetry minimum) is seen. With this we can identify the 3p-6T$_{1u}$ antibonding hybrid (AB) and the 6T$_{1u}$ C$_{60}$ molecular orbital.
The asymmetry-curve in Figure \ref{F4}b) is also not flat below B and above AB. This is likely related to the fact that the photoemission cross sections of all other molecular orbitals are affected by the endohedral unit as seen in the different HOMO:HOMO-1 intensity ratios.

In order to quantify the difference between Ar@C$_{60}$ and C$_{60}$ we subtract a 4$^{th}$ order polynomial background from the asymmetry curve in Figure \ref{F4} b) and reconstruct the corresponding difference between the Ar@C$_{60}$ and C$_{60}$ spectrum.
In Figure \ref{F4}c) this difference shows a splitting $\Delta$ between B and AB of 1.6$\pm$0.2 eV, which is close to the calculated value of 1.47 eV. 
As expected, the hybridizing 6T$_{1u}$ molecular orbital of C$_{60}$ shows up with negative values in the intensity difference between Ar@C$_{60}$ and C$_{60}$.
It lies 0.7 eV below AB, or 0.9 eV above B.
This suggests that AB has more 6T$_{1u}$ character and correspondingly B more Ar 3p character.
If 6T$_{1u}$ would lie in the middle between B and AB, no big difference between the intensity of B and AB would be expected.
The ratio between the B and AB intensity depends on the position in the hybridisation gap. 
Together with the fact that the He I$\alpha$ photoemission cross section is larger for an Ar 3p electron than for a C 2p electron this is consistent with the observation that B has a stronger cross section than AB.

Finally we would like to discuss the intensities, i.e. photoemission cross sections of the different molecular orbitals. 
For the 10 HOMO electrons of C$_{60}$ the experimental photoemission cross section at 21 eV photon energy is 100 Mb and 50 Mb for gas phase \cite{kor05} and condensed C$_{60}$ \cite{kor06}, respectively. 
Comparison of these cross sections with that of atomic C 2p (1.5 Mb/$e^-$) \cite{yeh85}) suggest for the molecule a C 2p cross section enhancement of a factor  7 to 3. 
The data shown in Figure \ref{F4} allow a comparison of the Ar@C$_{60}$ hybrid orbital cross section which turns out to be  0.44 $\pm 0.05$ times that of the HOMO, and is close to the value of 38 Mb for the Ar 3p level \cite{yeh85}.
This corresponds to the values as expected from the semiclassical result and thus we have no indication of a giant cross section enhancement in low energy photoemission of Ar in solid Ar@C$_{60}$, as it was proposed for the gas phase \cite{mad07}. 

In conclusion it is shown that in Ar@C$_{60}$, the Ar 3p and the C$_{60}$ 6T$_{1u}$ orbital strongly hybridize. This coupling between the endohedral unit and the carbon cage establishes Ar@C$_{60}$ as a noble gas compound, though imposes no enhancement of the photoemission cross section.

Technical support by Martin Kl\"ockner is gratefully acknowledged.
The project was financially support by the Swiss National Science Foundation.

\end{document}